\UseRawInputEncoding
\documentclass[pra,twocolumn,amsmath, superscriptaddress,notitlepage,showpacs]{revtex4-1}
\usepackage{amsmath,amsfonts,amssymb,amstext,amscd,amsthm,dsfont,bbm,hyperref,natbib,braket}
\usepackage[T1]{fontenc}
\usepackage{physics}
\usepackage{color}
\usepackage{soul,xcolor}
\usepackage{gensymb}
\usepackage{subcaption}
\hypersetup{
    colorlinks = true,
    citecolor=blue,
    linkcolor = red,
    anchorcolor = red,
    citecolor = blue,
    filecolor = red,
    pagecolor = red,
    urlcolor = blue,
    }
\usepackage[normalem]{ulem}
\usepackage{graphicx}
\usepackage{bbold}
\usepackage{braket}
\usepackage{placeins}

\newtheorem{theorem}{Theorem}
\newtheorem{lemma}{Lemma}

\begin{document}

\title{Initial entanglement, entangling unitaries, and completely positive maps}
	\author{Vinayak Jagadish}
	\email{jagadishv@ukzn.ac.za}
	\affiliation{Quantum Research Group, School  of Chemistry and Physics,
		University of KwaZulu-Natal, Durban 4001, South Africa}\affiliation{ National
		Institute  for Theoretical  Physics  (NITheP), KwaZulu-Natal,  South
		Africa}
			\author{R. Srikanth}
	\affiliation{Poornaprajna Institute of Scientific Research,
		Bangalore- 560 080, India}
	\author{Francesco Petruccione}
	\affiliation{Quantum Research Group, School  of Chemistry and Physics,
		University of KwaZulu-Natal, Durban 4001, South Africa}\affiliation{ National
		Institute  for Theoretical  Physics  (NITheP), KwaZulu-Natal,  South
		Africa}

\begin{abstract} 
The problem of conditions on the initial correlations between the system and the environment that lead to completely positive (CP) or not-completely positive (NCP) maps has been studied by various authors. Two lines of study may be discerned: one concerned with families of initial correlations that induce CP dynamics under the application of an arbitrary joint unitary on the system and environment; the other concerned with specific initial states that may be highly entangled. Here we study the latter problem, and highlight the interplay between the initial correlations and the unitary applied. In particular, for almost any initial entangled state,  one can furnish infinitely many joint unitaries that generate CP dynamics on the system. Restricting to the case of initial, pure entangled states, we obtain the scaling of the dimension of the set of these unitaries and show that it is of zero measure in the set of all possible interaction unitaries.
\end{abstract}
\maketitle  
\section{Introduction}
\label{intro}
The dynamics of a system $S$ is completely positive (CP) if $S$ begins initially in a product state with its environment $E$, irrespective of the interaction~\cite{petruccione}. However, if there are initial $S$-$E$ correlations, then the conditions under which they lead to positive or not-completely positive (NCP) maps has been studied extensively over the last decade and is still an active area of research~\cite{simmons_completely_1981,raggio_remarks_1982,simmons_another_1982,pechukas_reduced_1994,alicki_comment_1995,pechukas_pechukas_1995,jordan_dynamics_2004,carteret_dynamics_2008,cuffaro_debate_2013}.  In the presence of initial correlations, it was shown~\cite{stelmachovic_2001} that identical initial reduced states can lead to different evolutions under the same system-bath interaction, with the initial separable state leading to a CP map, but an entangled state leading to a NCP map. It was pointed out that the question of CPness depended not only on the initial correlation, but on the joint dynamics, and in particular that even with initial correlations, a factorizable joint dynamics will always lead to CP reduced dynamics~\cite{salgado_comment_2002}. This result was generalized to arbitrary initial correlations in~\cite{hayashi2003kraus}. It was argued that the system dynamics is CP if and only if the initial correlations lack quantum discord~\cite{shabani_vanishing_2009}, but the ``only if'' part was subsequently weakened~\cite{brodutch_vanishing_2013,shabani_erratum:_2016}.

Within the construct of direct-sum decomposition of state space, the set of initial states which include both separable and entangled states  for which the reduced dynamics is CP is addressed~\cite{liu_completely_2014}.  It was also pointed out that the dynamics is CP for the set of states if all quantum degrees of freedom are owned only by the system, and the classical degrees of freedom are of local accessibility to the system~\cite{lu_structure_2016}.
 It was also shown~\cite{buscemi_complete_2014} that initial correlations that lead to a violation of the data processing inequality correspond to NCP dynamics along with proving that system-bath correlations that can be obtained by steering a third (``reference'') system lead to CP dynamics under arbitrary joint unitaries, generalizing instances of the kind obtained earlier~\cite{shabani_erratum:_2016,brodutch_vanishing_2013}. Intuitively, this result may be understood as saying that bipartite correlations that are sufficiently weak as to be derivable by measuring the reference system in a sufficiently strongly correlated tripartite system allow CP dynamics of the system of interest. 

The above works can be classified in broadly two ways: works that investigate (sufficiently weak) initial correlations that entail CP dynamics under arbitrary joint unitaries, and those that study arbitrary initial correlations of specific states and their interplay with corresponding joint unitaries that would lead to CP dynamics. It is fair to say that while the former has been relatively well studied, the latter not so much. Accordingly, here, our focus will not be on generic initial states that generate CP dynamics under all system-environment unitaries, but instead to systematically point out how the interplay of the initial state and the joint dynamics determines CPness of the reduced dynamics. We thereby point out that even given initial entanglement, there are in general an infinite number of entangling unitaries that can lead to CP dynamics of the subsystem. 

The plan of this article is as follows. In Sec. \ref{entangunit}, we motivate the result by giving an example of an entangling unitary acting on an entangled state leading to CP dynamics. We then discuss a caveat and state the result in Sec. \ref{caveat}. The measure of CP inducing unitaries are addressed in Sec. \ref{measure}.  We then conclude in Sec. \ref{conclusion}.

\section{Entangling unitaries acting on entangled states}
\label{entangunit}
As mentioned in Sec. \ref{intro}, vanishing quantum discord  is not necessary for CP dynamics.  We now provide a specific illustration of this idea.
Consider the following entangled state 
\begin{equation}
\ket{\Psi} \equiv \frac{1}{\sqrt{2}}\left[
 (1+\imath) \cos (\theta )\ket{00}  + \sin (\theta)(\imath\ket{10} + \ket{11})\right],
 \label{eq:psi}
\end{equation}
where the system ($S$) is taken to be the second qubit and the first qubit is treated to be the environment ($E$). To this state, we apply $\sqrt{\rm CNOT} \equiv \ket0\bra0 \otimes \mathbbm{1} + \ket1\bra1 \otimes \sqrt{\sigma_x}$, ($\sigma_x$ being the Pauli matrix) which is an entangling unitary, with 
$$
\sqrt{\sigma_x} \equiv
\frac{1}{2}
\left(\begin{array}{cc}
 1+\imath& 1-\imath\\
 1-\imath & 1+\imath \\
\end{array}\right).
$$
At first blush, one might expect that acting such an entangling operation on an entangled state will lead to NCP reduced dynamics. But, looking at the the corresponding dynamical (Choi) matrix~\cite{sudarshan_stochastic_1961,Quanta77}, one finds that it turns out to be CP. Specifically, its eigenvalues, $\lambda_{\pm}$ are as follows 
\begin{equation}
\lambda_{\pm}=1\pm\frac{\sqrt{7+ 8 \cos (4 \theta )+ \cos (8 \theta )}}{3+\cos (4 \theta )},
\label{eigdynmat}
 \end{equation}
which can be clearly seen to be always positive. This example therefore shows that a total entangling unitary acting on an entangled state can admit CP dynamics.
There is a simpler way to understand why the action of $\sqrt{\rm CNOT}$ generates a CP dynamics here. It turns out that the state $\ket{\Psi}$ in Eq. (\ref{eq:psi}) can be expressed as the action of the same entangling operator on a product state: specifically, $\ket{\Psi} = \sqrt{\rm CNOT}(\cos\theta\ket0 + \sin\theta\ket1)\ket0$. Therefore, $\sqrt{\rm CNOT}\ket{\Psi} = {\rm CNOT}(\cos\theta\ket0 + \sin\theta\ket1)\ket0 = \cos\theta\ket{00} + \sin\theta\ket{11}$. 

In other words, the action of the entangling operation on $\ket{\Psi}$ in Eq. (\ref{eq:psi}) can be seen as an intermediate map of a CP-divisible channel acting on the initial state $(\cos\theta\ket0 + \sin\theta\ket1)\ket0$. But it is well recognized that the intermediate map of a CP-divisible channel-- in this the action of $\sqrt{\rm CNOT}$ on state $\ket{\Psi}$ is necesarily CP \cite{rivas_entanglement_2010}.  One should note that Eq. (\ref{eq:psi}) refers only to a particular example of an entangled state. However, the eigenvalues Eq. (\ref{eigdynmat}) are evaluated for the intermediate map, and not for a specific instance of a point to point transformation. This exercise alerts us to a class of entangling operations on entangled states, that will lead to CP dynamics. This idea provides a broad response to the belief sometimes held, as mentioned in Sec. \ref{intro}, that initial entangled correlations necessarily lead to NCP dynamics. This can be used to construct examples to show that non-vanishing quantum discord does not imply NCPness, and that there can be (infinitely many, as we shall indicate later) joint unitaries that lead to CP dynamics.  An important caveat is due in this context, which is discussed below.
 
\section{Issue of maximal entanglement}
\label{caveat}
Suppose as above, we apply CNOT unitary successively to the product state $(\cos\theta\ket0 + \sin\theta\ket1)\ket0$, with ${\rm CNOT} \equiv \ket0\bra0 \otimes \mathbbm{1} + \ket1\bra1 \otimes \sigma_x$, which has the property that ${\rm CNOT}^2 = \mathbbm{1}$.  We represent the map in the form acting on the density matrix expressed as a column vector and call it the $\mathcal{A}$ matrix following~\cite{sudarshan_stochastic_1961,Quanta77}.
Thus, the $\mathcal{A}$ matrix corresponding to the first application of CNOT is $\mathcal{A}_1$, and that corresponding to the second one is $\mathcal{A}_2$. We have $\mathcal{A}_1\cdot \mathcal{A}_2 = \mathbbm{1}_4$, and $\mathcal{A}_2$ can be evaluated to be
\begin{eqnarray}
\mathcal{A}_2 &=& (\mathcal{A}_1)^{-1} \nonumber \\
&=& \sec(2\theta)\left(
\begin{array}{cccc}
\cos ^2(\theta ) & 0 & 0 & -\sin ^2(\theta ) \\
0 & \cos ^2(\theta ) & -\sin ^2(\theta ) & 0 \\
0 & -\sin ^2(\theta ) & \cos ^2(\theta ) & 0 \\
-\sin ^2(\theta ) & 0 & 0 & \cos ^2(\theta ) \\
\end{array}
\right),\nonumber
\label{eq:a2}
\end{eqnarray}
which upon reshuffling yields the corresponding dynamical matrix (Choi) matrix~\cite{sudarshan_stochastic_1961,Quanta77},
\begin{equation}
\mathcal{B}_2 = \sec(2\theta)\left(
\begin{array}{cccc}
\cos ^2(\theta ) & 0 & 0 &  \cos ^2(\theta ) \\
0 & -\sin ^2(\theta )& -\sin ^2(\theta ) & 0 \\
0 & -\sin ^2(\theta ) & -\sin ^2(\theta ) & 0 \\
\cos ^2(\theta )  & 0 & 0 & \cos ^2(\theta ) \\
\end{array}
\right),
\label{eq:b2}
\end{equation}
whose non-vanishing eigenvalues are $-2 \sin ^2(\theta )\sec(2\theta)$ and $2\cos ^2(\theta)\sec(2\theta)$. 
Clearly, one of the two eigenvalues is negative for any choice of $\theta \ne 0, n\pi$ ($n$ an integer), showing that the dynamics $\mathcal{A}_2$, corresponding to the second action of CNOT is NCP~\cite{jagadish_measure2_2019} for such choices.  The disentanglement manifests for example as an increase in system state purity and hence in the separation of two suitable initial states of the system, leading to NCPness.

\begin{lemma}
	Given an initial state $\ket{\Phi}_{SE}$ that is not maximally entangled and a unitary $U(t) \equiv e^{-\imath H t}$ that acts on $\ket{\Phi}_{SE}$, the reduced dynamics of the system $S$ is CP if there exists a product state $\ket{\psi}_S \otimes \ket{\phi}_E$ such that $\ket{\Phi}_{SE} = e^{-\imath H s}\ket{\psi}_S \otimes \ket{\phi}_E$ for $s \ge 0$.
	\label{lem:if}
\end{lemma}
\textbf{Proof:}  Suppose $\ket{\Phi}_{SE} = e^{-\imath H s}\ket{\psi}_S \otimes \ket{\phi}_E$. Then, $U(t)\ket{\Phi}_{SE} = e^{-\imath H (s+t)}\ket{\psi}_S \otimes \ket{\phi}_E$ describes the joint state at time $t$, and it is known that if the initial state is a product, then the evolution in the interval $[-s,t]$ is  CP-divisible. In turn, CP-divisibility implies that intermediate dynamics of the system during the interval $[0,t]$ is CP ~\cite{rivas_entanglement_2010,hall2010}. The restriction of non-maximality is evident in the CNOT example above, where we find that the further evolution of a maximally entangled state can lead to disentanglement and hence NCP intermediate dynamics. 

\hfill $\blacksquare$

\begin{figure}
	\includegraphics[width=8cm]{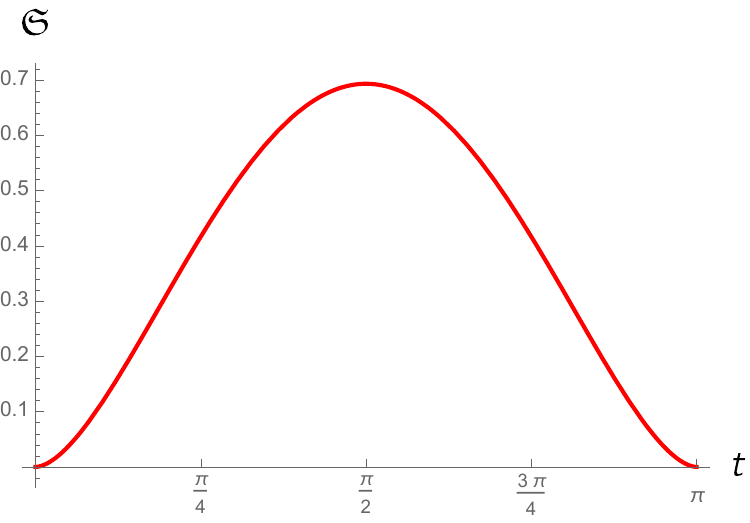}
	\captionsetup{singlelinecheck = false, format= hang, justification=raggedright, font=footnotesize,labelsep=space}
	\caption{(Color online) Entanglement of entropy $\mathfrak{S}$ for the dynamics described by Eq. (\ref{cnotme}). As a function of time, $t$, $\mathfrak{S}$ increases till $\pi/2$, and then falls. The corresponding decay rate of the master equation is given in FIG. \ref{fig:decay}.}
	\label{fig:entropy}
\end{figure}

 As a matter of semantic clarification, let us mention that by ``initial correlation'', we mean the correlation between the system and the environment at a fiducial instant of time designated $t=0$.  The argument that initial correlations should involve only product states is not only mathematically more restrictive, but is potentially unverifiable.
The unitary corresponding to the Hamiltonian $\ket0\bra0 \otimes \mathbbm{1} + \ket1\bra1 \otimes \sigma_x$ realizes the CNOT at $t=\pi/2$, when the second qubit is started in the state $\cos(\theta)\ket0 + \sin(\theta)\ket1$. The differential form of the map (for the choice of $\theta=\pi/4$) associated with the reduced dynamics of the first system under this interaction can be evaluated to
\begin{equation}
\frac{d\rho}{dt} = -\frac{\imath}{2}[\sigma_x, \rho] + \gamma(\sigma_x \rho \sigma_x - \rho),
\label{cnotme}
\end{equation}
with the decay rate $\displaystyle\gamma = \frac{\sin (2t)}{2(1+ \cos (2 t))}$.
The decay rate becomes negative after $t=\pi/2$ which is indicative of the fact that the map is CP-indivisible afterwards. If the two qubits start off in the product state $\frac{1}{\sqrt 2}(\ket0 +\ket1)_E \ket0_S$, one can see that entanglement is generated between $S$ and $E$ qubits from the plot of von-Neumann entropy $\mathfrak{S}$ with time, which is evaluated to be
\begin{equation}
\mathfrak{S}= \frac{1}{2}\Big(-(1+\cos (t)) \log [\cos ^2(\frac{t}{2})] + (\cos (t)-1) \log [\sin ^2(\frac{t}{2})]\Big),
\label{eq:entropy}
\end{equation}
plotted in Fig. \ref{fig:entropy}.
\begin{figure}
\captionsetup{singlelinecheck = false, format= hang, justification=raggedright, font=footnotesize, labelsep=space}
	\includegraphics[width=8cm]{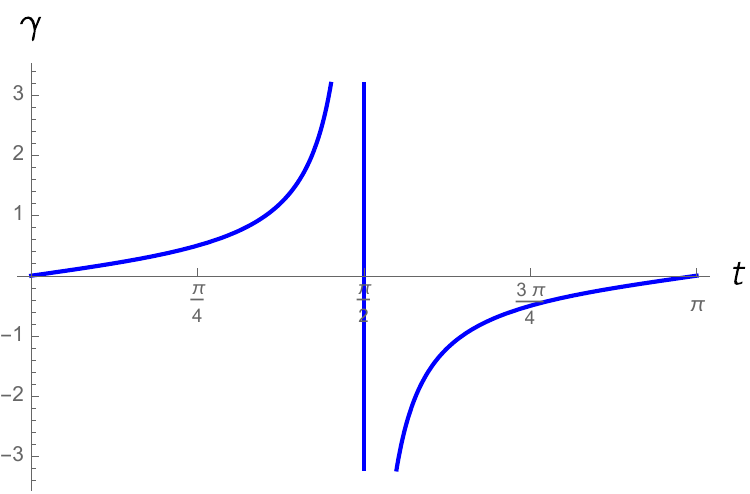}
	\caption{(Color online) Decay rate $\gamma$ as a function of time, $t$, for the dynamics described by Eq. (\ref{cnotme}).
		Referring to FIG. \ref{fig:entropy}, we observe that the dynamics is CP-divisible (positive decay rate) when the entanglement is non-zero and increasing, while the negative decay rate corresponds to the region of disentanglement.}
	\label{fig:decay}
\end{figure}
Two points are worth noting in reference to Figs. \ref{fig:entropy} and \ref{fig:decay}: (a) First is that the stationarity of the environment is not necessary for the CP-divisibility of the system dynamics. In particular, we note here that the entanglement is non-zero, and indeed increasing, in the range $t \in [0,\pi/2)$, while the decay rate remains positive, indicating CP-divisibility (b) The same interaction, after the attainment of maximal entanglement at $t=\pi/2$, induces CP-indivisible dynamics. In other words, even though the evolution of the system can be considered as starting with a product state, the further evolution past the state of maximal entanglement will lead to disentanglement, and thereby NCP dynamics. This gives a concrete instance of the exception made in the state of Lemma \ref{lem:if} with regard to maximally entangled states. 
 Note that similar conclusions hold for arbitrary $\theta$, other than $\theta = \pi/4$. 

\section{Measure of the set of unitaries inducing CP dynamics}
\label{measure}
The preceding surprising result evokes the question of the probability that, given an entangled state $\ket{\phi}_{SE}$, applying a random joint unitary will produce a CP reduced dynamics. Now, let us take the unitary to be $\sqrt{\rm CPHASE} \equiv \ket0\bra0 \otimes \mathbbm{1} + \ket1\bra1 \otimes \sqrt{\sigma_z}$ acting on the state $\ket{\Psi}_{SE}$ in Eq. (\ref{eq:psi}) and look at the reduced dynamics.

The corresponding dynamical matrix $\mathcal{B}$ turns out to be NCP. As an example, $\mathcal{B}$ for $\theta= \pi/4$ is
\begin{equation}
\mathcal{B}^{\pi/4} = \frac{1}{4}\left(
\begin{array}{cccc}
 4 & 0 & 1-\imath& 2-2\imath \\
 0 & 0 & 0 & -1+\imath \\
 1+\imath & 0 & 0 & 0 \\
 2+2\imath& -1-\imath & 0 & 4 \\
\end{array}
\right),
\end{equation}
with the eigenspectrum \{-0.0703, -0.2362, 0.5291, 1.7774\}, the two negative eigenvalues clearly indicating that the map is NCP. The important point to be noted is that the same entangled state Eq. (\ref{eq:psi}) when acted upon by two different unitaries made the reduced dynamics CP in the case of $\sqrt{\rm CNOT}$  and NCP for the case of $\sqrt{\rm CPHASE}$. 

 Per Lemma \ref{lem:if}, the backward evolution using the Hamiltonian $H_\phi \equiv \ket{0}\bra{0}\otimes \mathbbm{1} + \ket{1}\bra{1}\otimes \sigma_z$ that generates the C-Phase gate, will at no time lead to a product state. This may be explicitly checked by finding that the application of the evolution $e^{\imath H_\phi t}$ to the state in Eq. (\ref{eq:psi}) leads to a non-vanishing entanglement in the system $SE$. In fact, the reduced state of the system is 

\begin{equation}
\frac{1}{4}\left(
\begin{array}{cc}
 1 & - \imath e^{-\imath t} \\
\imath e^{\imath t} & 3 \\
\end{array}
\right),
\end{equation}
 with a constant von Neumann entropy of $\frac{1}{4} [5-2 \sqrt{2} \coth ^{-1}(\sqrt{2})]$, showing that the entanglement is a constant.

Despite the above result, one can show that given any initial non-maximally entangled state $\ket{\phi}_{SE}$,  there are infinitely many joint unitaries that, acting on a given entangled state $\ket{\Psi_{SE}}$ produce a CP dynamics of the system $S$ for a finite time which leads to the following result. 
\begin{theorem}
Given any pure non-maximally entangled state $\ket{\Phi}_{SE}$ of a system $S$ of dimension $d_S$ and its environment $E$ of dimension $d_E$, there are infinitely many entangling unitaries that induce CP dynamics on $S$. (a) The dimension $D$ of this set  of unitaries scales as $D = \mathcal{O}[(d_S)^{-2}]$ for a sufficiently large environment; (b) The set is of zero measure in the set of all interaction unitaries.
\label{lem:measure}
\end{theorem} 
\textbf{Proof of part (a) }  For any given (possibly entangled) state $\ket{\Phi}_{SE}$, there is a product state $\ket{\phi}_S\otimes\ket{\phi^+}_E$ and an entangling Hamiltonian $H$, such that we can construct a unitary $U_{SE}^{(1)}(t) \equiv e^{-\imath H t}$ and $\ket{\Phi}_{SE} = U_{SE}^{(1)}(t)\ket{\phi}_S\otimes\ket{\phi^+}_E$. Per Lemma \ref{lem:if}, there is an entangling unitary $U^{(2)}_{SE} \equiv  e^{-\imath H s}$, for sufficiently small time parameter $s$, such that $U^{(2)}_{SE}$ acting on $\ket{\Phi}_{SE}$ induces a CP dynamics on $S$ since it can be considered as an intermediate map of the system's CP-divisible evolution under $U^{(2)}_{SE}\ket{\Phi}_{SE} = e^{-\imath H(t+s)} \ket{\phi}_S\otimes\ket{\phi^+}_E$.

Now consider any other product state  $\ket{\psi}_S\otimes\ket{\psi^+}_E$ in the same subspace. There necessarily exists an entangling unitary $U_{SE}^\prime$ such that $\ket{\Phi}_{SE} = U_{SE}^\prime\ket{\psi}_S\otimes\ket{\psi^+}_E$, and we define  Hamiltonian $H^\prime$ such that $U_{SE}^\prime =  e^{-\imath H^\prime t}$. Then, one can write down another entangling unitary $U^{(2\prime)}_{SE} \equiv e^{-\imath H^\prime s}$, corresponding to the pre-initial product state $\ket{\psi}_S\otimes\ket{\psi^+}_E$, that would also induce a CP dynamics in the system $S$ starting in the same initial joint state $\ket{\Phi}_{SE}$. Therefore, given an entangling interaction unitary $U^{(2)}_{SE}$ that induces CP dynamics, we can produce another one corresponding to another pre-initial product state.

Extending the above reasoning, every product state can be used to construct an entangling unitary of the kind $U^{(2)}_{SE}$ that induces CP system dynamics acting on the initial state $\ket{\Phi}_{SE}$. The pre-initial product states can be put into a one-to-one correspondence with local unitaries $U_{\rm loc}$, which are elements of the group $SU(d_S) \times SU(d_E)$. This is because the set of all product states is equivalent to this group acting on a fiducial product state. On the other hand, an arbitrary interaction unitary is an element of the group $SU(d_S d_E)$. 

The ratio of the number of parameters required to characterize such a unitary versus an arbitrary joint unitary is, then, simply the ratio between the number of parameters required to specific an arbitrary local unitary versus an arbitrary joint unitary, namely, $$\frac{|SU(d_S) \times SU(d_E)|}{|SU(d_S d_E)|} \approx \frac{d_S^2 + d_E^2}{d_S^2d_E^2},$$ which tends to $(d_S)^{-2}$ in the limit that $d_E \gg d_S$.  It is possible that, for a given, specific initial state, other joint unitaries can be found that, acting on a given entangled state, induce CP dynamics on the system, hence the above count may be considered as a lower bound on the dimension of the set of such unitary operators. 

\textbf{Proof of part (b) }: 
The above result naturally leads to the following observation. For finite dimensions $d_S$ and $d_E$, the measure of pure product states is zero in the set of all pure states, i.e., an arbitrary bipartite state is almost always entangled. The elements of the set $\mathcal{S}_P$ of all pure product states can be put into a one-to-one correspondence with local unitaries in the set $SU(d_S) \times SU(d_E)$, by considering each element of $\mathcal{S}_P$ as being obtained by the action of a local unitary on a fiducial product state. By a similar reasoning, the elements of the set $\mathcal{S}_2^\ast$ of all pure bipartite states here can be put into one-to-one correspondence with the set $SU(d_S d_E)$ of all bipartite unitaries. Thus, the set of product states is a lower-dimensional subset of the set of all states, having a zero measure. 

Consider a given initial state $\ket{\Psi}_{SE}$ subjected to an arbitrary entangling Hamiltonian $H_{SE}$, with eigenstates $\ket{\lambda_j}$ and corresponding eigenvalues $\lambda_j$. Letting $\ket{\Psi}_{SE} \equiv \sum_j \alpha_j \ket{\lambda_j}$, the subsequent evolution of the state has the form $\ket{\Psi^\prime}_{SE}(t) \equiv \sum_j \alpha_je^{i\lambda_j t} \ket{\lambda_j}$, with the evolution picking up relative phase factors. The set $\eta \equiv \{\ket{\Psi^\prime}_{SE}(t) , t\ge 0\}$ of all future states of the system is evidently also a zero-measure set in $\mathcal{S}_2^\ast$. Therefore, the measure of the intersection of these two zero-measure sets, namely $\eta \cap \mathcal{S}_P$, is zero. In other words, the state obtained by applying an arbitrary unitary to the given initial entangled state is almost never a product. Since unitaries are time-symmetric, this also means that an arbitrary Hamiltonian evolution applied in reverse time to a given entangled state will almost never produce a product state. Thus an arbitrary unitary applied to $\ket{\Phi}_{SE}$ almost always leads to NCP dynamics (per the usual expectation). \hfill $\blacksquare$

We now consider the question of whether we can systematically identify entangling unitaries that, acting on a given entangled state, lead to CP dynamics. Quite generally, we may remark the following: suppose, following Theorem \ref{lem:measure} that the initial entangled state $\ket{\Psi}_{SE} \equiv U_{SE}(\ket{\phi_1}_S \otimes \ket{\phi_2}_E)$ is such that  $U^\prime_{SE}\ket{\Psi}_{SE}$ induces CP dynamics on the system $S$. This means that  $U_{SE}^\prime$ leads to the intermediate map of the CP-divisible map under the total evolution generated by $U_{SE}^\prime U_{SE}$. Let $U_L$ be a local unitary with the property that $U_L^\prime \equiv U_{SE}^\dagger U_L U_{SE}$ is also a local unitary. Then,  $(U^\prime_{SE}U_L)\ket{\Psi}_{SE}$ will also correspond to CP dynamics of $S$. 
	
To show this note that $U^\prime_{SE}U_L \ket{\Psi}_{SE} = U_{SE}^\prime U_L U_{SE}(\ket{\phi_1}_S \otimes \ket{\phi_2}) = U_{SE}^\prime U_{SE} U_L^\prime (\ket{\phi_1}_S \otimes \ket{\phi_2})$, i.e., leading to a  CP-divisible evolution evolution given by the action of the joint unitary $U_{SE}^\prime U_{SE}$ on the pre-initial product state $U_L^\prime (\ket{\phi_1}_S \otimes \ket{\phi_2}_E)$. By Lemma \ref{lem:if}, this will lead to CP dynamics on $S$.

As examples of this idea, suppose $U_{SE} = U_{SE}^\prime =\sqrt{\rm CNOT}$ as in Example 1. One notes that $U_L := \sigma_z \otimes \sigma_x$ preserves its locality when transformed under $U_{SE}^\prime$. Then, it follows that the action of $ U_{SE}^\prime U_L \equiv \sqrt{\rm CNOT}(\sigma_z \otimes \sigma_x)$ will correspond to CP dynamics of system $S$. Yet another example would be $\sqrt{\rm CNOT}(\sigma_z \otimes \sqrt[\leftroot{-1}\uproot{2}\scriptstyle n]{\sigma_x})$, with $n$ being any integer. On the other hand, we find that $U_{SE} U_L U_{SE}^\dagger \equiv \sqrt{\rm CNOT}(\sigma_x \otimes \sigma_x)\sqrt{\rm CNOT}^\dagger$ is not a local unitary, and correspondingly, $ U_{SE}^\prime U_L \equiv \sqrt{\rm CNOT}(\sigma_x \otimes \sigma_x)$ will correspond to NCP dynamics of system $S$. More generally, if $U_{SE} = \ket{0}\bra{0}\otimes \mathbbm{1} + \ket{1}\bra{1}\otimes U$, then a local operator $U_L \equiv \sigma_z\otimes V$, such that $[U,V]=0$ or $\{U,V\}=0$, can be used to augment $U_{SE}^\prime$ to $U_{SE}^\prime U_L$ while retaining the CP dynamics of the system $S$.

 \section{Conclusions}
 \label{conclusion}
The problem of initial entanglement and CP maps still poses new questions, not least the issue of whether the concept of NCP maps is an artifact of an ill-defined evolution map \cite{schmid_why_2019}. The structure and characterization of initial correlations and NCP has lately been studied by a number of authors, e.g. \cite{liu_completely_2014,lu_structure_2016,brodutch_vanishing_2013,shabani_erratum:_2016,buscemi_complete_2014}. In this article, aided with simple examples, we have shown the interplay of the initial correlations and the joint dynamics of the system and environment in deciding the CPness of the reduced dynamics of the system of interest. In particular, for any initial entangled state (barring maximal entanglement) of the system and environment,  one can furnish a two-body unitary that generates CP dynamics on the system, and for any given interaction unitary  one can furnish an entangled state such that the action of the latter unitary generates CP reduced dynamics on the system. We showed that in fact there are in general an infinite number of such entangling unitaries that can act on a given (non-maximal) entangled initial state, but still lead to CP dynamics of the system. Specifically, we obtained the scaling of the dimension of the set of these unitaries, and point out that their measure in the set of all possible interaction unitaries is zero. It is an interesting question how the above result generalizes for mixed states, which we leave open. The general description of entangling operators that may be applied to a given entangled state leading to CP dynamics should be an interesting problem of future study.

\section{Acknowledgements}
The work of V.J. and F.P. is based upon research supported by the South African Research Chair Initiative of the Department of Science and Innovation and National Research Foundation (NRF) (Grant UID: 64812).  R.S.   acknowledges the support of Department of Science and Technology (DST), India, Grant No.: MTR/2019/001516.

\bibliography{EntanglementCP.bib}

\end{document}